\documentclass[conference,a4paper]{IEEEtran}
\usepackage{cite}
\usepackage{amsmath,amssymb,amsfonts}
\usepackage{graphicx}
\usepackage{textcomp}
\usepackage{xcolor}
\usepackage{booktabs}
\usepackage{url}
\usepackage{balance}

\def\BibTeX{{\rm B\kern-.05em{\sc i\kern-.025em b}\kern-.08em
    T\kern-.1667em\lower.7ex\hbox{E}\kern-.125emX}}

\begin{document}

\title{Band-Selective Microwave Cavity Optimization Using Differentiable FDTD: Gradient-Guided Search Versus Structured Random Search}

\author{
\IEEEauthorblockN{Hasan Yiğit\IEEEauthorrefmark{1}, Kutlu Karayahşi\IEEEauthorrefmark{2}}
\IEEEauthorblockA{\IEEEauthorrefmark{1}Department of Software Engineering, Muğla Sıtkı Koçman University, Türkiye\\
Email: hasanyigit@mu.edu.tr}
\IEEEauthorblockA{\IEEEauthorrefmark{2}Department of Electrical and Electronics Engineering, Muğla Sıtkı Koçman University, Türkiye\\
Email: kutlukarayahsi@mu.edu.tr}
}

\maketitle

\begin{abstract}

We compare gradient-based inverse design with structured random search for dielectric-loaded microwave cavities using an in-house JAX-based differentiable FDTD solver. The benchmark fixes material fraction, filtered density representation, initialization, band-energy objective, and measured selection wall time. Each selected design is evaluated in a separate 8000-step FDTD simulation. Across four target bands, three prescribed material fractions, and six seeds, gradient optimization achieves a higher in-band spectral-energy fraction in all 72 paired comparisons. The mean paired improvement is 0.356, with a 95\% paired bootstrap interval of 0.339–0.374 and an exact two-sided sign-flip p value of 0.03125. Gradient, cavity-mode, CFL, material-fraction, and raw-output integrity checks pass before inference. Under matched material and computational budgets, the results support an advantage for gradient-based optimization in this cavity-design benchmark.
\end{abstract}

\vskip0.5\baselineskip
\section{Introduction}

Inverse design is useful when the relation between a material layout and an
electromagnetic response has no convenient closed form. Microwave cavities are
an example: their spectra depend on many coupled material variables, and every
design evaluation requires a full-wave solve~\cite{molesky2018}. The engineering task here is
spectral response shaping: use a fixed amount of dielectric to direct a larger
share of the measured cavity response into a prescribed microwave band.
This is a benchmark design problem for cavity-resonator filters, RF packages and
enclosures, and resonant sensing structures, before port, loss, and manufacturing
constraints are introduced. The bands studied here (48--104\,GHz) lie in the
millimeter-wave spectrum used by 5G/6G radios, radar, and short-range sensing
links, where package- and cavity-level spectral control directly affects
antenna efficiency, shielding, and interference.
Adjoint methods offer an efficient approach to such high-dimensional problems by obtaining a full design
gradient from a small number of field solves~\cite{aage2017}.

Automatic differentiation provides another route to those gradients. It
differentiates the time-stepping program directly and avoids a separate,
hand-derived adjoint~\cite{bradbury2018}. Recent differentiable FDTD work shows
that this is practical for electromagnetic inverse design~\cite{tang2023,
mahlau2026,kim2026}. Differentiability alone, however, does not establish a
fair or physically reliable optimization comparison.

That gap motivates this study. A credible comparison must verify the gradient,
the solver, the physical material constraint, and the computational budget. If
these controls are absent, an apparent gain can instead arise from extra
dielectric loading, a different representation, a longer time window, or more
computation. Closed cavities add a further challenge: nearby discrete modes can
make a point-frequency objective reward the wrong resonance. We therefore use
the fraction of spectral energy inside a target band and use the same 8000-step
window for selection and validation.

Our research question is direct: under matched material, representation,
initialization, objective, and selection wall time, does gradient optimization
achieve a higher in-band spectral-energy fraction $R_B$ than structured random
search when each selected design is evaluated in a separate FDTD simulation?
We test four bands, three prescribed material fractions, and six seeds with a
differentiable JAX-based FDTD solver.

Under these matched conditions, we hypothesize that gradient optimization
yields a higher post-optimization $R_B$ than structured random search.

The paper makes three contributions. First, it defines a material- and
time-matched benchmark for gradient and structured-random cavity design.
Second, it verifies the gradient, solver, and post-optimization metric before testing
performance. Third, it quantifies the engineering outcome---the fraction of
the measured response placed in the requested band---across all 72 paired
evaluation outcomes with a pre-specified seed-level inference procedure.

\section{Methodology}

The idea is to treat the entire pipeline---filtered density projection,
8000-step FDTD time stepping, and the band-energy metric---as a single
differentiable program, so that one reverse-mode pass through the time
stepping returns the gradient of the band fraction with respect to every
design logit, without a hand-derived adjoint. Both compared methods search
the same physical design space with the same objective and differ only in
whether this gradient is used; the budget they are given is measured in
wall time, not in candidate counts.

\subsection{Cavity and FDTD Formulation}
\label{sec:formulation}
\begin{figure}[t]
\centering
\includegraphics[width=\columnwidth]{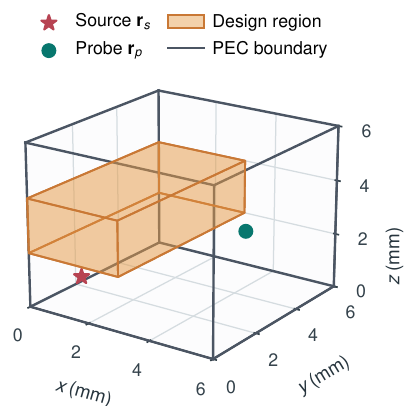}
\caption{Simulation domain: $6\times6\times6\,\text{mm}$ PEC cavity, source and probe cells, and the dielectric design region (orange) at one wall.}
\label{fig:geometry}
\end{figure}

All calculations use our in-house, version-controlled differentiable FDTD
implementation written in JAX. The domain is a
$6\times6\times6\,\text{mm}$ perfectly electrically
conducting (PEC) cavity. It has 30 Yee-cell intervals per direction and 31
field nodes, with $\Delta=0.2\,\text{mm}$. The physical distance between
opposite PEC faces is therefore exactly $6\,\text{mm}$. The source and probe
are at $(1,1,1)\,\text{mm}$ and $(5,3,3)\,\text{mm}$, respectively
(Fig.~\ref{fig:geometry}). The probe records $E_z(t)$. The domain is
\begin{equation}
\Omega = [0,L]^3 \subset \mathbb{R}^3, \quad L = 6\,\text{mm}, \quad \partial\Omega \ \text{PEC},
\label{eq:domain}
\end{equation}
Fields obey Faraday's and Amp\`ere's laws,
\begin{equation}
\frac{\partial \mathbf{H}}{\partial t} = -\frac{1}{\mu}\nabla\times\mathbf{E}, \qquad \frac{\partial \mathbf{E}}{\partial t} = \frac{1}{\varepsilon}\nabla\times\mathbf{H},
\label{eq:maxwell}
\end{equation}
and are advanced on a staggered Yee grid with the standard leapfrog update~\cite{yee1966,warnick2011,houle2020} in normalized units ($\varepsilon_0=\mu_0=1$, impedance-scaled $\mathbf{H}$),
\begin{align}
\mathbf{H}^{n+\frac12} &= \mathbf{H}^{n-\frac12} - \Delta t\,\nabla_h^{+}\times\mathbf{E}^{n}, \\
\mathbf{E}^{n+1} &= \mathbf{E}^{n} + \frac{\Delta t}{\varepsilon_r}\,\nabla_h^{-}\times\mathbf{H}^{n+\frac12},
\label{eq:yee}
\end{align}
with $E_{\mathrm{tangential}}=0$ on PEC faces. The normalized selection model
and the separate post-optimization FDTD evaluation use float64. With outward
normal $\mathbf{n}$, the boundary and initial conditions are explicitly
\begin{align}
\mathbf{n}\times\mathbf{E}\rvert_{\partial\Omega} &= \mathbf{0}, \\
\mathbf{E}(\mathbf{r},0) &= \mathbf{0}, \qquad \mathbf{r}\in\Omega, \\
\mathbf{H}(\mathbf{r},-\tfrac12\Delta t) &= \mathbf{0}, \qquad \mathbf{r}\in\Omega.
\label{eq:bcic}
\end{align}
A broadband Gaussian pulse is then injected into $E_z$ at the source cell.
The time step is $365.7\,\text{fs}$, or $0.9493$ of the three-dimensional CFL
limit.

\begin{figure*}[t]
\centering
\includegraphics[width=0.96\textwidth]{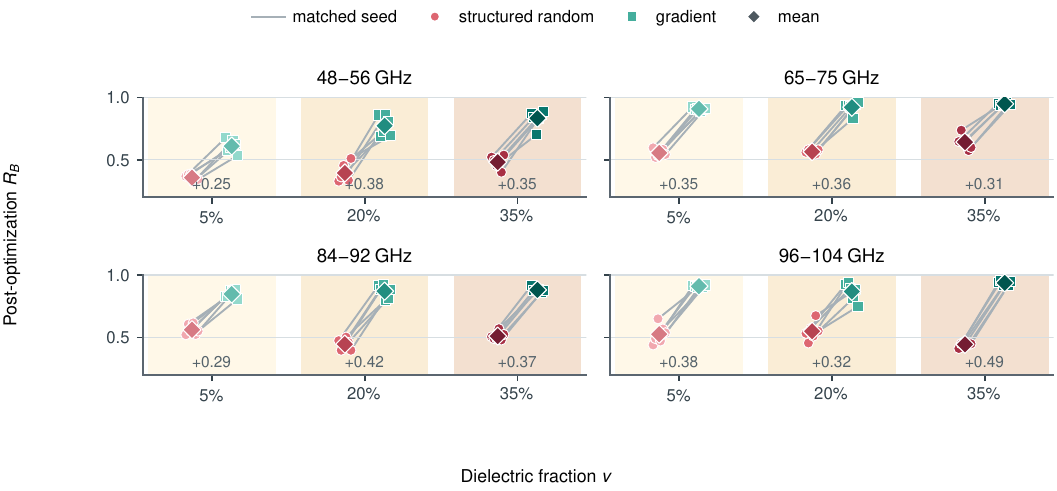}
\caption{Post-optimization $R_B$ for $v=5\%,20\%,35\%$. Lines pair seeds;
circles/squares mark structured-random/gradient seed values, and diamonds mark
means. Rose/teal distinguish the methods; shade darkens with $v$.}
\label{fig:effects}
\end{figure*}

\begin{figure*}[t]
\centering
\includegraphics[width=0.96\textwidth]{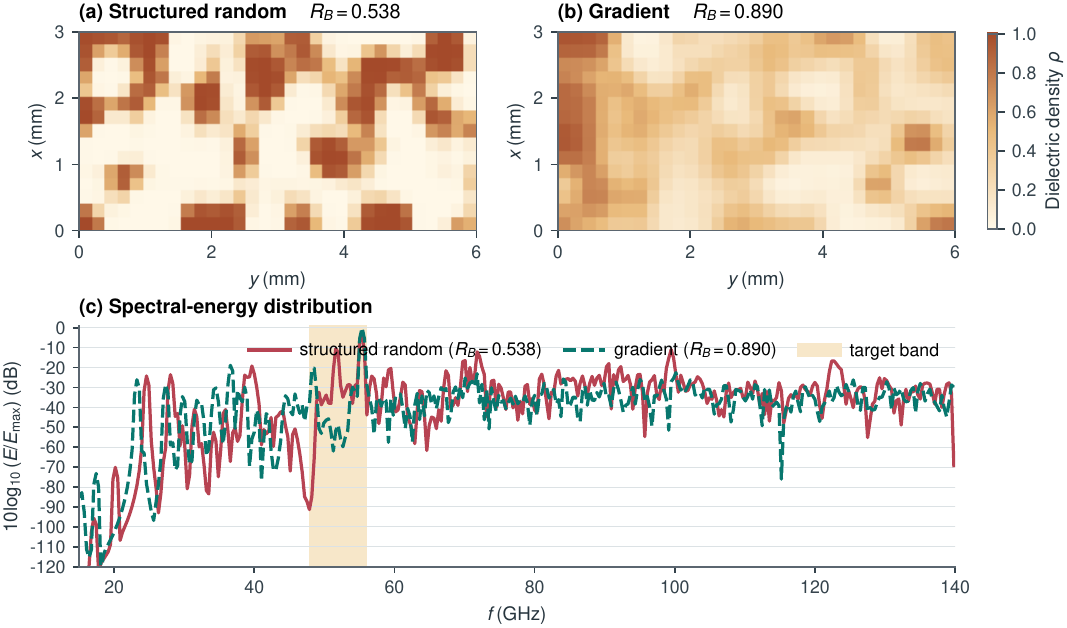}
\caption{Automatically selected median-effect matched pair from the recorded
post-optimization outputs (48--56\,GHz, $v=35\%$, seed 5). Top: physical dielectric
density in the wall design region for (a) structured random search ($R_B=0.538$)
and (b) gradient optimization ($R_B=0.890$). (c) Spectral-energy distributions
computed from the corresponding recorded 8000-step FDTD traces, on a common
0-dB reference; the shaded interval is the target band. The pair difference is
$\Delta R_B=+0.352$.}
\label{fig:design-pair}
\end{figure*}

\subsection{Design Parameterization}
\label{sec:design}
A $15\times31$ raw-logit field parameterizes a dielectric region at one cavity
wall and is extruded through ten grid planes. Both methods use the same
two-cell cone filter~\cite{bourdin2001}. A differentiable scalar shift is then solved so that the
mean physical density equals the prescribed fraction $v\in\{0.05,0.20,0.35\}$
exactly. The isotropic permittivity is
\begin{equation}
\varepsilon_r=1+29\rho, \qquad 0\leq\rho\leq1,
\label{eq:eps}
\end{equation}
inside the design region and one elsewhere. Thus the comparison matches the
amount of dielectric after filtering and projection, not merely a raw-logit
penalty.

\subsection{Band-Selective Objective}
\label{sec:objective}
For a target band $[f_{\mathrm{lo}},f_{\mathrm{hi}}]$, both selection and
reporting use the Parseval rFFT band-energy
fraction~\cite{oppenheim2010,virtanen2020}
\begin{equation}
R_B = \frac{\sum_{k:\,f_k \in [f_{\text{lo}},f_{\text{hi}}]} c_k\,|U[k]|^2}{\sum_{k} c_k\,|U[k]|^2}, \quad U = \mathrm{rFFT}(w \odot u),
\label{eq:rb}
\end{equation}
where $w$ is a Hann window~\cite{harris1978} and $c_k$ is the one-sided Parseval
weight ($c_k=1$ at DC and, for even-length traces, at Nyquist; $c_k=2$
otherwise)~\cite{oppenheim2010,virtanen2020}. The selection loss is
$-\log(R_B+10^{-8})$. This objective is differentiable and $R_B\in[0,1]$.
Selection and validation now share the same frozen 8000-step metric, so their
agreement is a consistency requirement rather than independent evidence; its
substance is the recomputation over production tasks. Across all 144
post-optimization tasks, the normalized selection scores and the separate
SI-FDTD evaluations of $R_B$ agree with Spearman correlation one and a
maximum absolute difference of $6.1\times10^{-15}$.

\subsection{Optimization}
Gradient optimization uses Adam (learning rate 0.3) for 100 iterations. Each
iteration uses an 8000-step normalized rollout. The final design is then
evaluated by a separate 8000-step FDTD simulation; only that
post-optimization value of $R_B$ enters the results.

\subsection{Structured Random Baseline and Technical Gates}
The structured-random baseline is intentionally used as a simple
gradient-free reference to isolate the benefit of gradient information under
the same measured computational budget.
Structured random search evaluates filtered, volume-projected candidates with
the same loss. Its first candidate is the gradient method's initial logit
field, and its remaining candidates follow a fixed, seed-specific structured
stream. It runs until the measured gradient-selection wall time is reached.
Consequently, candidate count is not the budget; measured selection time is.
All six seeds are repeated for every band--fraction pair.
Across the 144 optimized task instances, recorded task wall time totaled
$106.8$ h: $97.7$ h for matched selection and $9.1$ h for post-optimization
evaluations; fixed warm-ups and four controls added $1.0$ h.

Before inference, the pre-specified evaluation design checks the numerical
solver, gradient, and experiment records. Table~\ref{tab:checks} reports the
decisive checks.
The final analysis also recomputes $R_B$ from every stored post-optimization trace and
verifies every output hash.

\begin{table}[t]
\centering
\caption{Pre-inference technical checks.}
\label{tab:checks}
\footnotesize
\begin{tabular}{@{}lp{0.46\linewidth}@{}}
\toprule
Check & Observed value \\
\midrule
3-D CFL fraction of limit & $0.9493$ (limit: $\leq0.95$) \\
Selected empty-cavity modes & max. relative error $2.41\times10^{-3}$ \\
Gradient finite difference & coordinate $2.24\times10^{-8}$; directional max. $7.55\times10^{-9}$ \\
Material-fraction error & max. $2.78\times10^{-16}$ (144 designs) \\
Random/gradient selection time & $1.00003$--$1.00179$ (allowed: $1.00$--$1.01$) \\
Raw-output audit & 144/144 hashes and recomputed $R_B$ values matched \\
\bottomrule
\end{tabular}
\end{table}

\section{Results}
\label{sec:results}

\begin{table}[t]
\centering
\caption{Post-optimization cavity response. $R_B$: probe spectral-energy
fraction in the target band (higher is better); volume: prescribed dielectric
fraction. Values: six-seed paired means; $\Delta$: gradient minus structured
random; Wins: seeds where gradient is higher.}
\label{tab:results}
\scriptsize
\begin{tabular*}{0.95\columnwidth}{@{\extracolsep{\fill}}ccrrrr@{}}
\toprule
Band & Volume & Gradient & \shortstack{Structured\\random} & $\Delta$ & Wins \\
\midrule
48--56 & 5\%  & 0.610 & 0.358 & +0.252 & 6/6 \\
48--56 & 20\% & 0.773 & 0.395 & +0.378 & 6/6 \\
48--56 & 35\% & 0.833 & 0.481 & +0.351 & 6/6 \\
\addlinespace[1pt]
65--75 & 5\%  & 0.907 & 0.555 & +0.352 & 6/6 \\
65--75 & 20\% & 0.921 & 0.565 & +0.356 & 6/6 \\
65--75 & 35\% & 0.948 & 0.642 & +0.306 & 6/6 \\
\addlinespace[1pt]
84--92 & 5\%  & 0.847 & 0.561 & +0.286 & 6/6 \\
84--92 & 20\% & 0.870 & 0.446 & +0.424 & 6/6 \\
84--92 & 35\% & 0.880 & 0.510 & +0.370 & 6/6 \\
\addlinespace[1pt]
96--104 & 5\%  & 0.910 & 0.526 & +0.384 & 6/6 \\
96--104 & 20\% & 0.867 & 0.549 & +0.318 & 6/6 \\
96--104 & 35\% & 0.936 & 0.444 & +0.492 & 6/6 \\
\bottomrule
\end{tabular*}
\end{table}

Gradient optimization exceeded structured random search in all 72 paired
comparisons (Table~\ref{tab:results}). The mean paired difference was 0.356.
In short, under identical material budgets and measured selection wall time,
gradient-based design outperformed structured random search in every one of
the 72 paired tasks.
In engineering terms, under the same allowed dielectric fraction, gradient
optimization directed a larger fraction of the measured cavity response into
the requested frequency band. This is a band-selection (spectral-response
shaping) result, not an insertion-loss or $Q$-factor measurement.
For the primary analysis, the pre-specified unit of inference is the seed; we
therefore use six seed-level averages across all 12 band--material conditions.
All six were
positive; the exact two-sided sign-flip test gave $p=0.03125$, and the 95\%
paired bootstrap interval was $[0.339,0.374]$. Figure~\ref{fig:effects} shows
all 72 matched seed-level post-optimization values as well as the condition means.
Figure~\ref{fig:design-pair} makes the corresponding physical layouts and
recorded spectra concrete for one automatically selected representative pair:
different density layouts yield different spectral allocations, and the
gradient layout has the larger target-band fraction for this median-effect pair.
The twelve condition-level effects are descriptive: their individual exact
tests become non-significant after Holm adjustment ($p_{\mathrm{Holm}}=0.375$
for each), so the inferential claim is the pre-specified global effect rather
than twelve separate claims.

\section{Discussion and Conclusion}
\label{sec:discussion}

Under this matched experimental design, gradient optimization has a
statistically supported advantage over structured random search for
band-selective cavity design. This conclusion rests on separate FDTD evaluation
traces, not optimization-time scores, and it remains after material amount and
measured selection time are matched.

This study shows that, for the present cavity-design problem, access to
gradients changes not merely the optimization score but the resulting
electromagnetic response. When dielectric loading and selection time are held
constant, gradient-based design achieves a larger $R_B$, meaning that a greater
fraction of the cavity response is placed in the requested microwave band.

The central outcome is therefore a more reliable route from a spectral
requirement to a physically interpretable dielectric layout. This establishes a
practical computational foundation for gradient-based band-selective cavity optimization and raises the next
question: how far can the same design logic be carried when the spectral
requirement is tied to a specific microwave component or system?

\section*{Acknowledgment}
The authors directed and verified the technical decisions, simulations, and numerical claims. AI-assisted tools supported code development, visualization, and manuscript preparation under author supervision. All quantitative figures were generated directly from the recorded simulation outputs; they are not AI-generated imagery.

\balance


\begin{thebibliography}{00}

\bibitem{molesky2018} S. Molesky, Z. Lin, A. Y. Piggott, W. Jin, J. Vuckovi\'c, and A. W. Rodriguez, ``Inverse design in nanophotonics,'' \emph{Nature Photonics}, vol. 12, pp. 659--670, 2018, doi: 10.1038/s41566-018-0246-9.

\bibitem{aage2017} N. Aage and V. E. Johansen, ``Topology optimization of microwave waveguide filters,'' \emph{Int. J. Numer. Methods Eng.}, vol. 112, no. 3, pp. 283--300, 2017, doi: 10.1002/nme.5551.

\bibitem{bradbury2018} J. Bradbury \emph{et al.}, ``JAX: composable transformations of Python+NumPy programs,'' version 0.3.13, 2018. [Online]. Available: http://github.com/jax-ml/jax

\bibitem{tang2023} R. J. Tang \emph{et al.}, ``Time reversal differentiation of FDTD for photonic inverse design,'' \emph{ACS Photonics}, vol. 10, no. 12, pp. 4140--4150, 2023, doi: 10.1021/acsphotonics.3c00694.

\bibitem{mahlau2026} Y. Mahlau, F. Schubert, L. Berg, and B. Rosenhahn, ``FDTDX: High-performance open-source FDTD simulation with automatic differentiation,'' \emph{J. Open Source Softw.}, vol. 11, no. 117, 2026, doi: 10.21105/joss.08912.

\bibitem{kim2026} B. Kim, ``rfx: JAX-based differentiable 3D FDTD simulator for RF engineering,'' REMI Lab, Chungnam National University, 2026. [Online]. Available: https://github.com/bk-squared/rfx

\bibitem{yee1966} K. S. Yee, ``Numerical solution of initial boundary value problems involving Maxwell's equations in isotropic media,'' \emph{IEEE Trans. Antennas Propag.}, vol. AP-14, no. 3, pp. 302--307, May 1966, doi: 10.1109/TAP.1966.1138693.

\bibitem{warnick2011} K. F. Warnick, \emph{Numerical Methods for Engineering: An Introduction Using MATLAB and Computational Electromagnetics Examples}. Raleigh, NC, USA: SciTech Publishing, 2011, doi: 10.1049/SBEW049E.

\bibitem{houle2020} J. E. Houle and D. M. Sullivan, \emph{Electromagnetic Simulation Using the FDTD Method with Python}, 3rd~ed. Piscataway, NJ, USA: IEEE Press/Wiley, 2020, doi: 10.1002/9781119565826.

\bibitem{bourdin2001} B. Bourdin, ``Filters in topology optimization,'' \emph{Int. J. Numer. Methods Eng.}, vol. 50, no. 9, pp. 2143--2158, 2001, doi: 10.1002/nme.116.

\bibitem{oppenheim2010} A. V. Oppenheim and R. W. Schafer, \emph{Discrete-Time Signal Processing}, 3rd~ed., Prentice-Hall signal processing series. Upper Saddle River, NJ, USA: Pearson, 2010. [Online]. Available: https://books.google.com.tr/books?id=mYsoAQAAMAAJ

\bibitem{virtanen2020} P. Virtanen \emph{et al.}, ``SciPy 1.0: fundamental algorithms for scientific computing in Python,'' \emph{Nature Methods}, vol. 17, pp. 261--272, 2020, doi: 10.1038/s41592-019-0686-2.

\bibitem{harris1978} F. J. Harris, ``On the use of windows for harmonic analysis with the discrete Fourier transform,'' \emph{Proc. IEEE}, vol. 66, no. 1, pp. 51--83, 1978, doi: 10.1109/PROC.1978.10837.

\end{thebibliography}
\end{document}